\documentclass{raa}
\usepackage{graphicx,times,rotating}
\usepackage{graphicx}
\usepackage{amsmath}
\input{epsf.sty}
\input{psfig.sty}

\begin{document}

\title{Broad-band spectroscopy of the transient X-ray binary pulsar KS~1947+300 
during 2013 giant outburst: Detection of pulsating soft X-ray excess component}

\author{Prahlad Epili, Sachindra Naik and Gaurava K. Jaisawal}

\institute{Astronomy and Astrophysics Division, Physical Research Laboratory, Ahmedabad, India\\
              {\it prahlad@prl.res.in (PE); snaik@prl.res.in (SN); gaurava@prl.res.in (GKJ)}
}

\abstract {We present the results obtained from detailed timing 
and spectral studies of the Be/X-ray binary pulsar KS~1947+300 during 
its 2013 giant outburst. We used data from {\it Suzaku} observations 
of the pulsar at two epochs i.e. on 2013 October 22 (close to the peak 
of the outburst) and 2013 November 22. X-ray pulsations at $\sim$18.81~s 
were clearly detected in the light curves obtained from both  
observations. Pulse periods estimated during the outburst showed 
that the pulsar was spinning up. The pulse profile was found to be 
single-peaked up to $\sim$10~keV beyond which a sharp peak followed 
by a dip-like feature appeared at hard X-rays. The 
dip-like feature has been observed up to $\sim$70~keV. 
The 1-110~keV broad-band spectroscopy of both  
observations revealed that the best-fit model comprised of a 
partially absorbed Negative and Positive power law with EXponential 
cutoff (NPEX) continuum model along with a blackbody component for 
the soft X-ray excess and two Gaussian functions at 6.4 and 6.7~keV 
for emission lines. Both the lines were identified as emission from 
neutral and He-like iron atom. To fit the spectra, we 
included the previously reported cyclotron absorption line at 12.2 keV. 
From the spin-up rate, the magnetic field of the pulsar 
was estimated to be $\sim$1.2$\times$10$^{12}$~G and found to be 
comparable to that obtained from the detection of the cyclotron 
absorption feature. Pulse-phase resolved 
spectroscopy revealed the pulsating nature of the soft X-ray excess
component in phase with the continuum flux. This confirms that the 
accretion column and/or accretion stream are the most probable regions 
of pulsar's soft X-ray excess emission. The presence of the pulsating 
soft X-ray excess in phase with continuum emission may be the possible 
reason for not observing the dip at soft X-rays. 
}

\authorrunning{P. Epili, S. Naik \& G. K. Jaisawal}
\titlerunning{Suzaku observations of KS~1947+300}

\maketitle

\section{Introduction}

Be/X-ray binaries (BeXBs) are known to be the largest subclass ($\sim$60\%) of 
high-mass X-ray binaries (Caballero \& Wilms 2012). The majority of BeXBs consist of  
a neutron star as compact object and a Be star as optical companion. 
The optical companions in these binary systems are non-supergiant B-type stars
of luminosity class III-V that show emission lines at their 
evolutionary phase (Okazaki \& Negueruela 2001; Reig 2011). The neutron star 
in BeXBs accretes matter from the circumstellar disk of Be star usually at  
periastron passage. This abrupt accretion of huge amount of matter causes 
significant enhancement of the X-ray emission from the pulsating neutron star. 
These periodic enhancements of the X-ray intensity  
are known as Type~I X-ray outbursts (peak luminosity $\sim$ 10$^{35}$ - 10$^{37}$ 
ergs s$^{-1}$). The neutron star in these systems, however, occasionally 
shows rare X-ray outbursts (Type~II), lasting for several tens of days to a few
months during which the peak luminosity reaches up to 10$^{38}$ ergs s$^{-1}$. 
For a brief review of the properties of BeXBs, refer to Paul \& Naik (2011).

The transient Be/X-ray binary pulsar KS~1947+300 was discovered on 1989 June 8 
with the {\it Kvant/TTM} coded-mask X-ray spectrometer on {\it Mir} space station 
(Skinner 1989; Borozdin et al. 1990). The pulsar was first detected at a flux 
level of 70~mCrab that later decreased to $\sim$10~mCrab within 2 months 
of the detection. The spectra obtained from these observations were described 
by an absorbed power-law with photon-index of 1.72$\pm0.31$ (Borozdin et al. 
1990). In 1994 April, the X-ray pulsar GRO~J1948+32 was discovered close to the 
coordinates of KS~1947+300 by the Burst and Transient Source Experiment 
(BATSE) instrument on-board {\it Compton Gamma Ray Observatory (CGRO)} 
(Chakrabarty et al. 1995). The pulsation from this new source was found 
to be 18.7~s. Based on the spin period analysis, later Swank \& Morgan 
(2000) established that KS~1947+300 and GRO~J1948+32 
are the same source. A Be star of 14.2 visible magnitude at a distance of 
$\sim$10~kpc was discovered as the optical counterpart of the pulsar 
(Negueruela et al. 2003). 

KS~1947+300 was inactive from 1995 to 2000 without showing any major X-ray 
outburst. Subsequent {\it RXTE} observations showed that the source 
became active in October 2000 and went into a strong X-ray outburst (Levine \& 
Corbet 2000). The pulsar spectrum in the 2-80 keV range during the outburst was 
described with a Comptonization continuum model along with a blackbody component 
for the soft X-ray excess and a Gaussian component for the iron emission line at 6.5~keV 
(Galloway et al. 2004). The orbital parameters of the binary system were also 
estimated from observations during October and are reported 
in Galloway et al. (2004). A glitch in the pulsar frequency, generally 
seen in anomalous X-ray pulsars and radio pulsars and rare in accretion powered X-ray 
pulsars, was first detected in KS~1947+300 (Galloway et al. 2004). A low frequency 
quasi-periodic oscillation (QPO) at 20~mHz was detected at several occasions in  
declining phase of 2001 outburst with {\it RXTE} (James et al. 2010). Detection of 
low frequency QPOs and strong pulsations at low luminosity levels indirectly 
indicated that the magnetic field of the neutron star is $<10^{13}$~G 
as predicated from spin-up torque and luminosity correlation (James et al. 2010). Using 
{\it BeppoSAX} observations of 2001 X-ray outburst, Naik et al. (2006) described the 
broad-band pulsar spectrum in 0.1-100~keV with a Comptonization model along with a  
blackbody component at $\sim$0.6~keV and detected a weak 6.7~keV emission line from 
helium-like iron atoms. However, iron K$_\alpha$ emission line was absent in  
the spectrum during 2001 outburst.

A series of weak outbursts were observed from KS~1947+300 during the period of 
2002-2004. Among these outbursts, the strongest one was detected with {\it INTEGRAL} 
in April 2004. A high energy cutoff power-law model was used to describe the spectra 
obtained from {\it INTEGRAL/ISGRI} and {\it JEM-X} data during this outburst 
(Tsygankov et al. 2005). However, there was no signature of cyclotron 
absorption lines in the pulsar spectra. In October 2013, KS~1947+300 was detected 
in a giant outburst with a peak flux of 130~mCrab in 3-10~keV range (F{\"u}rst et 
al. 2014). Several {\it Swift/XRT} and three {\it NuSTAR} pointed observations were 
performed at different phases of the X-ray outburst. Combined spectra from 
{\it Swift/XRT} and {\it NuSTAR} in 0.8-79~keV range were described by a power 
law with an exponential cutoff continuum model along with a blackbody and an iron 
line component at 6.5~keV (F{\"u}rst et al. 2014).  A cyclotron absorption line 
at $\sim$12.2~keV was discovered in the pulsar spectra only during the second 
{\it NuSTAR} observation. The surface magnetic field of the pulsar was estimated 
to be $\sim$1.1$\times10^{12}$(1+z)~G. Though KS~1947+300 had gone through several 
major X-ray outbursts, cyclotron absorption line was not detected in the  
spectra obtained from earlier {\it RXTE}, {\it BeppoSAX} or {\it INTEGRAL} 
observations. 

As 2013 October outburst was a giant outburst, the pulsar was active for 
a few months during which it was observed with different X-ray observatories. 
We have used two {\it Suzaku} observations of the pulsar during this outburst 
to study its broad-band timing and spectral properties. The details on the 
observations, analysis, results and conclusions are presented in the following 
sections of the paper.

\section{Observations and  Data Analysis}

The fifth Japanese X-ray satellite, {\it Suzaku}, was launched on 2005, July 10 
 (Mitsuda et al. 2007). The X-ray 
imaging spectrometers (XISs; Koyama et al. 2007) and the hard X-ray detectors 
(HXDs; Takahashi et al. 2007) onboard {\it Suzaku} covered the 0.5-600 keV 
range. The XISs are CCD cameras located 
at the focal points of each of the four X-ray telescopes (XRTs; 
Serlemitsos et al. 2007). XISs (XIS-0, XIS-2 and XIS-3) 
are front-illuminated while XIS-1 is back-illuminated. The hard 
X-ray unit of {\it Suzaku} consists of two non-imaging detectors e.g. HXD/PIN 
and HXD/GSO. HXD/PIN consists of silicon diode detectors that work in 10-70 keV 
range and HXD/GSO consists of crystal scintillator detectors covering 40-600~keV 
range. The field of view of XIS is $17.8\arcsec\times17.8\arcsec$ in open window 
mode. HXD/GSO has same field of view of $34\arcsec\times34\arcsec$ as HXD/PIN up 
to 100~keV.

\begin{figure}
\centering
\includegraphics[height=3.2in, width=2.2in, angle=-90]{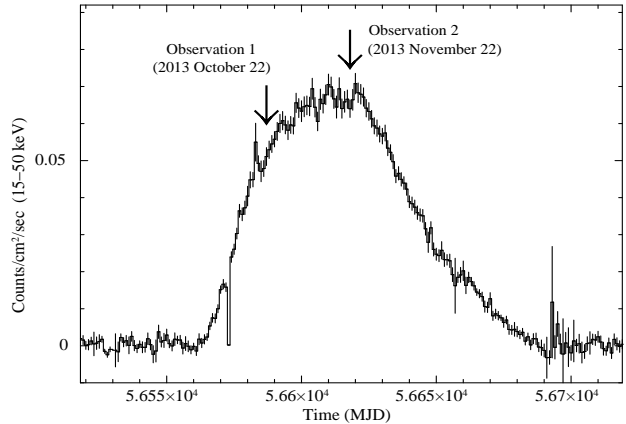}
\caption{{\it Swift/BAT} light curve of KS~1947+300 in the 15-50 keV energy range 
from 2013 August 14 (MJD 56518) to 2014 March 03 (MJD 56719). The arrow marks in 
the figure show the dates of {\it Suzaku} observations of the pulsar during the 
giant X-ray outburst.}
\label{sw}
\end{figure}

Target of opportunity (ToO) observations of KS~1947+300 were carried out with 
{\it Suzaku} at two epochs during its giant outburst in 2013 October-November. 
The first observation was performed on 2013 October 22, close to the peak of the
outburst for an effective exposure of $\sim$29~ks. The second 
observation was carried out on 2013 November 22-23 at the peak of the outburst 
for an effective exposure of $\sim$8~ks and 32~ks for XIS and HXD, respectively. 
The XIS detectors were operated in normal and burst clock mode with 2~s and 
0.5~s time resolutions for first and second observations, respectively. Both 
the observations were carried out in `XIS nominal' position. {\it Swift/BAT} 
monitoring light curve of the pulsar in 15-50 keV range covering the giant 
X-ray outburst is shown in Fig.~\ref{sw}. The arrows marks in the figure 
represent the date of {\it Suzaku} observations of the pulsar. In the present 
study, we used the publicly available data with Observation IDs: 908001010 and 
908001020 (hereafter Obs.1 and Obs.2).

\begin{table}
\centering
\caption{Pulse period history of KS~1947+300 during its 2013 giant outburst.}
\begin{tabular}{lllll}
\hline
\hline 
\\
 MJD         &Date           &Period (s)            &References \\
\\
\hline
\\
 56586.79    &2013-10-21     &18.80584(16) 	    &F{\"u}rst et al. (2014)  \\
 56587.22    &2013-10-22     &18.8088(1)            &present work \\
 56618.61    &2013-11-22     &18.7878(1)   	    &present work \\
 56618.91    &2013-11-22     &18.78399(7)  	    &F{\"u}rst et al. (2014)  \\
 56635.75    &2013-12-09     &18.77088(6)  	    &F{\"u}rst et al. (2014)  \\
 57053       &2015-01-31     &18.76255              &Finger et al. (2015)    \\
\hline
\hline
\end{tabular}
\label{pulse}
\end{table}

\subsection{Data reduction}

Unfiltered XIS and HXD event data were reprocessed by using {\it aepipeline} 
task in Heasoft (version 6.16) analysis package. Calibration database (CALDB) 
files released on 2012 November 06 (XIS) and  2011 September 13 (HXD) were 
used during data reprocessing. Cleaned event files generated after reprocessing 
were used in the present study. The {\it aebarycen} task of FTOOLS was applied 
on the cleaned event data to neutralize the effects of motions of the satellite 
and the earth around the Sun. We checked the effect of thermal flexing by 
applying the attitude correction S-lang script
\textit{aeattcor.sl}\footnote{http://space.mit.edu/ASC/software/suzaku/aeattcor.sl}  
on XISs data. After attitude correction, the XIS cleaned events were examined 
for possible photon pile-up by using the S-lang script 
\textit{pile\_estimate.sl}\footnote{http://space.mit.edu/ASC/software/suzaku/pile\_estimate.sl}.
During Obs.1, we detected a pile-up of $\sim$31\%, 21\% \& 33\% at the 
centers of XIS-0, XIS-1 and XIS-3 images, respectively. Therefore, an 
annulus region with inner and outer radii of 75\arcsec and 200\arcsec was 
chosen to reduce the pile-up below 4\%. As in case of Obs.1, we estimated 
photon pile-up for Obs.2 which was found to be $\sim$15\%, 12\% and 18\% at 
the centers of XIS-0, XIS-1 and XIS-3 images, respectively. An annulus region 
with inner and outer radii of 35\arcsec and 200\arcsec was considered for the 
pile-up correction in second observation. The light-curves and spectra of the 
pulsar were extracted from the XIS cleaned event data by applying the annulus 
regions in {\it XSELECT} package. Background light curves and spectra were 
accumulated from a source free region in the XIS image frame. Response and 
effective area files for XISs were created from the ``resp=yes'' task 
during the spectral extraction in {\it XSELECT}.  
HXD/PIN and HXD/GSO source light curves and spectra were extracted from the
cleaned event data by applying GTI selection in {\it XSELECT}. The PIN 
and GSO background light curves and spectra were generated in a similar manner 
from the simulated tuned non-X-ray background event data provided by the 
instrument team. The response file released on June 2011 was used for HXD/PIN 
for both the observations. However, GSO response and additional effective area 
files released on 2010 May 24 and 2010 May 26, respectively, were used 
while analyzing HXD/GSO data.

\begin{figure*}
\centering
\includegraphics[height=5.4in, width=3.2in, angle=-90]{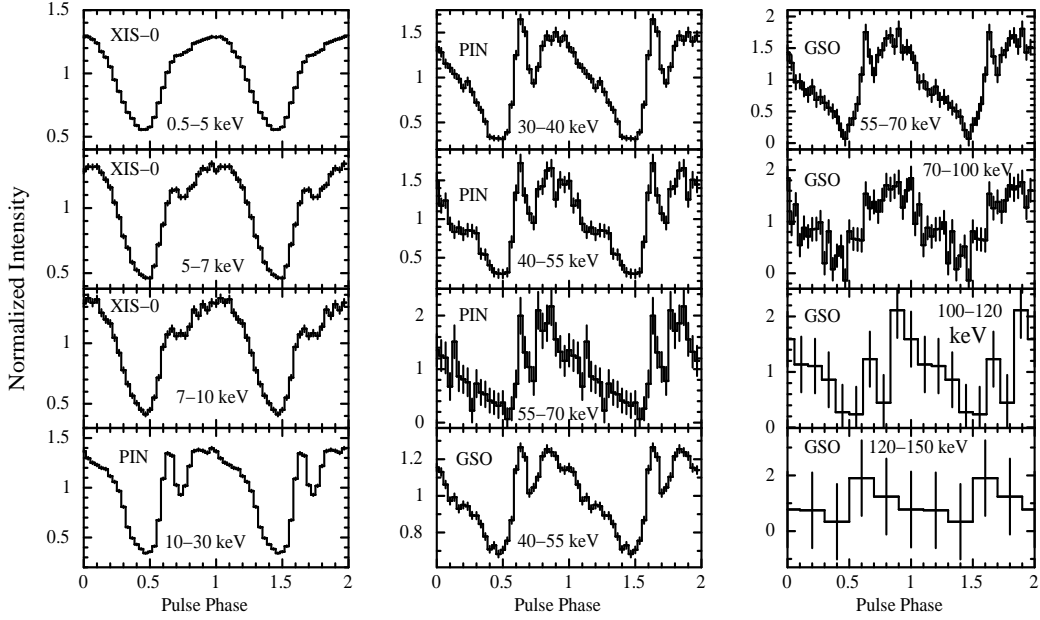}
\caption{Energy-resolved pulse profiles of KS~1947+300 obtained from XIS-0, HXD/PIN and 
HXD/GSO light curves at various energy ranges obtained from the first {\it Suzaku} observation 
of the pulsar on 2013 October 22. The error bars represent 1$\sigma$ uncertainties. 
Two pulses in each panel are shown for clarity.}
\label{er1}
\end{figure*}

\section{Timing Analysis}

Source and background light curves were extracted from the reprocessed 
and barycentric corrected XISs, PIN \& GSO event data with time resolutions of 
2~s, 1~s, 1~s for Obs.1 and 0.5~s, 1~s, 1~s for Obs.2, respectively. The 
$\chi^{2}$-maximization technique was applied to search for the periodicity 
in the background subtracted XIS and PIN light curves. Pulsations at periods of 
18.8088(1) and 18.7878(1)~s were detected in the light curves obtained from the 
first and second {\it Suzaku} observations, respectively. Though the 
observations were carried out within a time interval of one month, the decrease 
of the pulse period during later epoch suggests that the pulsar was 
spinning up. The pulse period history of KS~1947+300, obtained from {\it NuSTAR} 
and {\it Suzaku} observations during October-November 2013 outburst, is given 
in Table~\ref{pulse}. Decreasing values of the spin period with time confirmed 
that the pulsar was spinning-up during the outburst. Recent measurement of 
pulse period of KS~1947+300 in 2015 January-February outburst (see Table~\ref{pulse}) 
also indicated the long term spin-up trend in the pulsar. 

\begin{figure*}
\centering
\includegraphics[height=5.4in, width=3.2in, angle=-90]{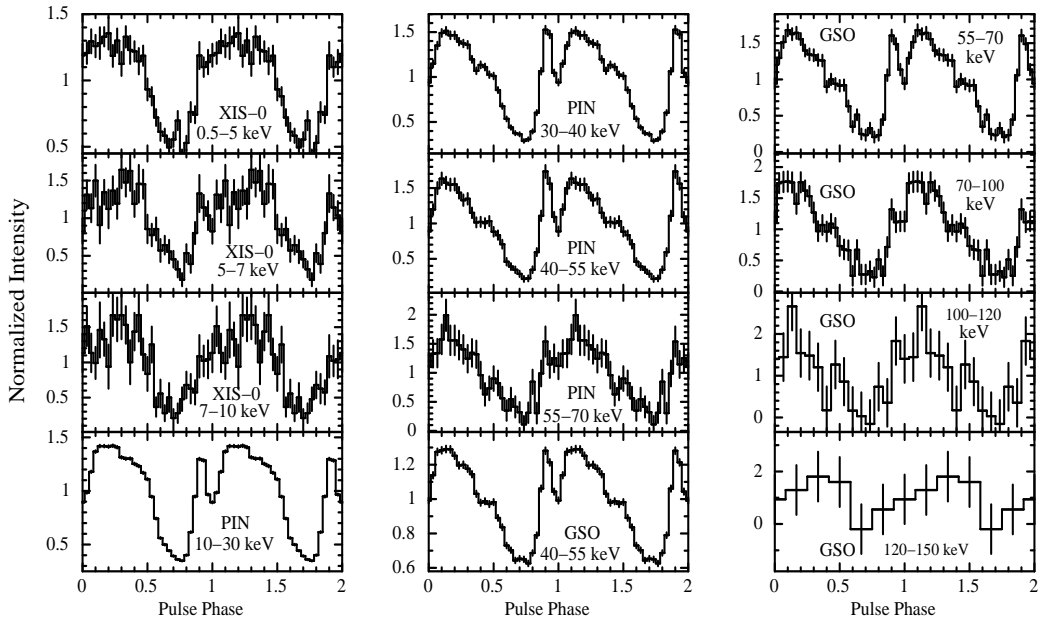}
\caption{Energy-resolved pulse profiles of KS~1947+300 obtained from XIS-0, HXD/PIN and 
HXD/GSO light curves at various energy ranges, during second {\it Suzaku} observation. The 
presence of dip-like feature can be clearly seen in the pulse profiles. The error bars 
represent 1$\sigma$ uncertainties. Two pulses in each panel are shown for clarity.}
\label{er2}
\end{figure*}

Pulse profiles of the pulsar in different energy bands were generated 
from the XIS, PIN and GSO light curves, obtained from both observations, 
are shown in Fig.~\ref{er1} \& \ref{er2}. During the first observation, the soft 
X-ray pulse profile 
(below 5~keV) was found to be smooth and single peaked. However, with 
increase in energy, a narrow dip-like feature appeared in the pulse profile 
and became prominent in 30-40~keV range. Beyond this energy, the depth of dip 
decreased and disappeared from the pulse profiles in 70-100~keV range. During 
second observation, the pulse profiles were found to show similar type of 
energy dependence as seen during first observation. X-ray pulsations in 
the light curves were detected up to $\sim$120~keV and $\sim$150~keV 
during first and second observations, respectively. Absorption dips in
the pulse profiles of Be/X-ray binary pulsars during X-ray outbursts are found 
to be prominent in soft X-rays and weak in hard X-ray bands (Paul \& Naik 2011 
and references therein; Naik et al. 2013; Naik \& Jaisawal 2015). However, in 
the case of {\it Suzaku} observations of KS~1947+300, the dip was found to be absent 
in the soft X-ray pulse profiles and prominent in the hard X-ray pulse profiles. It 
is, therefore, interesting to investigate the spectral properties of the pulsar 
to understand the causes of the absence/presence of absorption dip in  
soft/hard X-ray pulse profiles.

\section{Spectral Analysis}

\subsection{Pulse-phase averaged spectroscopy}

We performed phase-averaged spectroscopy of KS~1947+300 by using data 
from both {\it Suzaku} observations carried out during the giant outburst.
Data from XIS-0, XIS-1, XIS-3, PIN and GSO detectors were used in our
analysis. The procedure followed to extract source and background spectra 
was described in the earlier section. To improve statistics, we re-binned 
the source spectra obtained from XISs and PIN event data to have a minimum 
of 20 counts per energy channel. However, GSO spectra were grouped as 
suggested by instrument team. As in case of other bright X-ray sources 
where a systematic error of up to 3\% was added to XIS spectra (Cyg~X-1; 
Nowak et al. 2011), a systematic error of 1\% was added to XIS spectra 
of KS~1947+300 for the cross calibration issues between back and front 
illuminated CCDs. Simultaneous spectral fitting was carried out in 
$\sim$1-110~keV range for both observations by using {\it XSPEC} 
(version 12.8.2) package. During spectral fitting, data in 1.7-1.9 keV 
and 2.2-2.4~keV ranges were ignored due to the presence of known Si and 
Au edges in XISs spectra. All the model parameters were tied together 
except the values of relative normalization of detectors which were kept 
free during simultaneous spectral fitting.

We used a high-energy cut-off power law, a cut-off power law and the NPEX model to describe 
the pulsar continuum. We found that all three models can explain the  
continuum spectrum well. Along with the interstellar absorption, a blackbody 
component for the soft X-ray excess and a Gaussian function for iron emission
were required to fit the spectra. Though the broad-band spectral fitting 
yielded an emission line at $\sim$6.5~keV with a width of $\sim$0.2~keV, 
careful investigations of the residuals near the line energy confirmed the 
presence of two iron emission lines at 6.4 and 6.7~keV. Therefore, we added Gaussian
functions at 6.4 and 6.7~keV in our broad-band spectral fitting. We identified 
these lines as emission from neutral and He-like iron atoms. It was found 
that the addition of partial covering component to above continuum models improved 
the spectral fitting further with significant improvement in the $\chi^2$ values
($\Delta\chi^2$ $\ge$ 70). This component has been used to investigate the cause of 
absorption dips at certain phases of the pulse profiles of Be/X-ray binary pulsars 
(Paul \& Naik 2011). This component, therefore, was used to probe the cause of 
observed absorption dip in the pulse profiles of KS~1947+300. A cyclotron line at 
12.2~keV as detected from {\it NuSTAR} observations was also included in the spectral 
model. Since {\it Suzaku} data cannot constrain well the line region, in our analysis,
we fixed the cyclotron line parameters i.e. line energy at 12.2 keV, width at 2.5 keV 
and depth at 0.16 as obtained from {\it NuSTAR} observation (F{\"u}rst et al. 2014). Among the 
three continuum models, the partial covering NPEX model along with other spectral 
components was found to be the best-fit model for both {\it Suzaku} observations. 

\begin{figure}
\centering
\includegraphics[height=3.6in, width=2.6in, angle=-90]{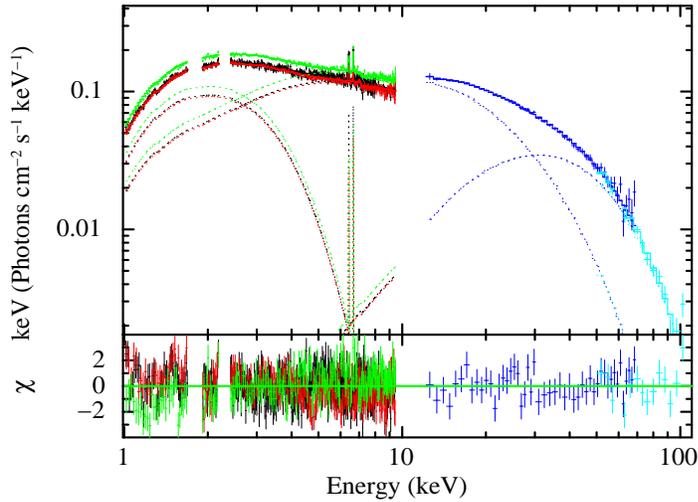}
\caption{Broad-band (1-110~keV energy range) spectrum of KS~1947+300 obtained with 
the XIS-0, XIS-1, XIS-3, PIN, and GSO detectors of first {\it Suzaku} observation 
during the 2013 October outburst along with the best-fit model comprising a partially 
absorbed NPEX continuum model, a blackbody component for soft X-ray excess, a Gaussian 
function for the iron emission line and fixed cyclotron absorption component. The
contributions of the residuals to the $\chi^2$ for each energy bin for the 
best-fit model are shown in the bottom panel.}
\label{spec1}
\end{figure}

\begin{figure}
\centering
\includegraphics[height=3.6in, width=2.6in, angle=-90]{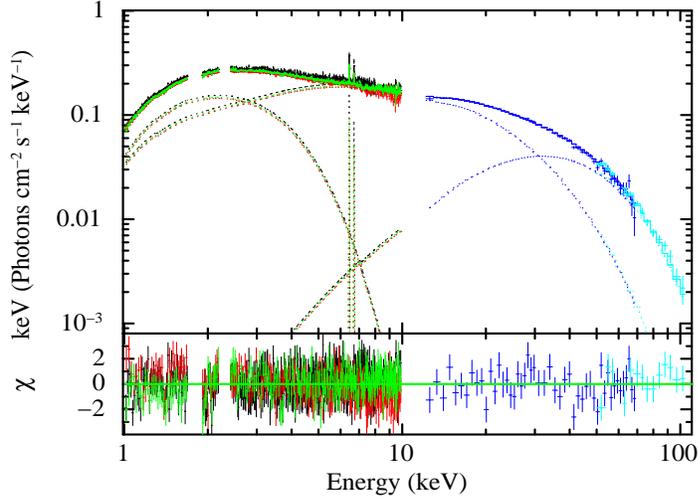}
\caption{1-110~keV energy spectrum of KS~1947+300 obtained with the XIS-0, XIS-1, XIS-3, 
PIN, and GSO detectors of second {\it Suzaku} observation during the 2013 October
outburst along with the best-fit model comprising a partially absorbed NPEX continuum 
model, a blackbody component for soft X-ray excess, a Gaussian function for the iron 
emission line and fixed cyclotron absorption component. The contributions of the residuals 
to the $\chi^2$ for each energy bin for the best-fit model are shown in the bottom 
panel.}
\label{spec2}
\end{figure}

Best-fitted spectral parameters obtained from all three models 
are given in Table~\ref{spec_par} for both  observations. The energy spectra 
for the partial covering NPEX continuum model along with residuals are shown in 
Figure~\ref{spec1} \& \ref{spec2} for first and second {\it Suzaku} observations, 
respectively. The values of additional absorption column density (N$_{H_2}$) were 
found to be significantly higher than the values of Galactic absorption column 
density (N$_{H_1}$) (Table~\ref{spec_par}). The pulsar spectrum was marginally 
hard at the peak of the outburst i.e. during second observation compared to the 
first observation. The soft excess component was found to be stronger during the 
second observation (peak of the outburst) with higher values of blackbody temperature 
and flux compared to that during the first observation.

\subsection{Pulse-phase resolved spectroscopy}

\begin{figure*}
\centering
\includegraphics[height=5.in, width=4.in, angle=-90]{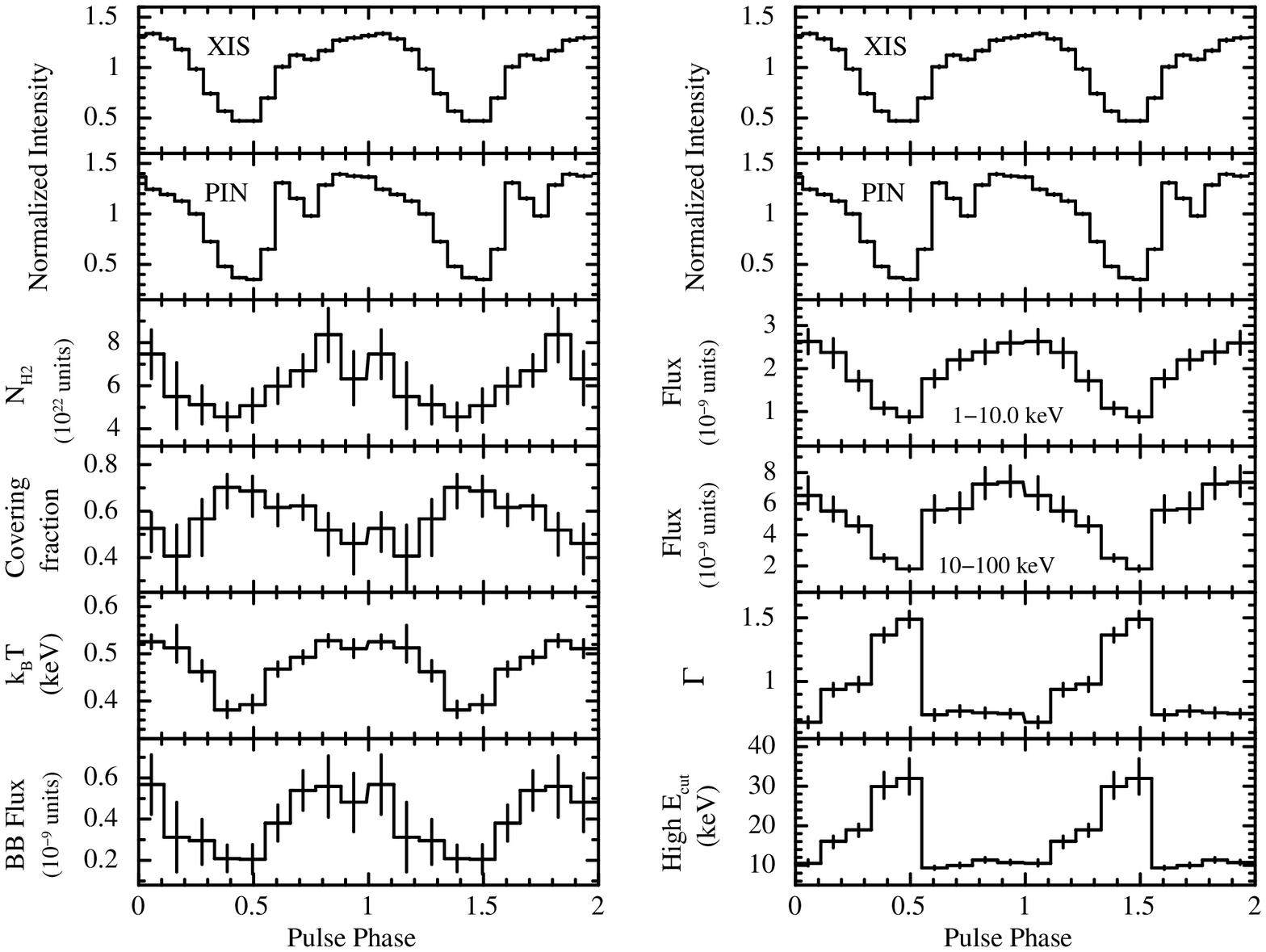}
\caption{Spectral parameters obtained from the phase-resolved spectroscopy 
of KS~1947+300 during first ${\it Suzaku}$ observation in 2013 October. 
The first and second panels in both sides show pulse profiles of the pulsar 
in 0.5-10~keV (XIS-0) and 10-70~keV (HXD/PIN) energy ranges. The values of 
$N_{H_2}$, covering fraction, blackbody temperature and blackbody flux
are shown in third, fourth, fifth and sixth  panels in left side of the figure, 
respectively. The  soft X-ray flux in 1-10 keV range, 
hard X-ray flux in 10-100 keV range,  photon index and high energy cutoff 
are shown in third, fourth, fifth and sixth panels in right side 
of the figure, respectively. The blackbody flux and source fluxes in 1-10 and 
10-100 keV are quoted in the units of $10^{-9}$ ergs cm$^{-2}$ sec$^{-1}$. The 
errors in the figure are estimated for 90\% confidence level.}
\label{pr1}
\end{figure*}

\begin{figure*}
\centering
\includegraphics[height=5.in, width=4.in, angle=-90]{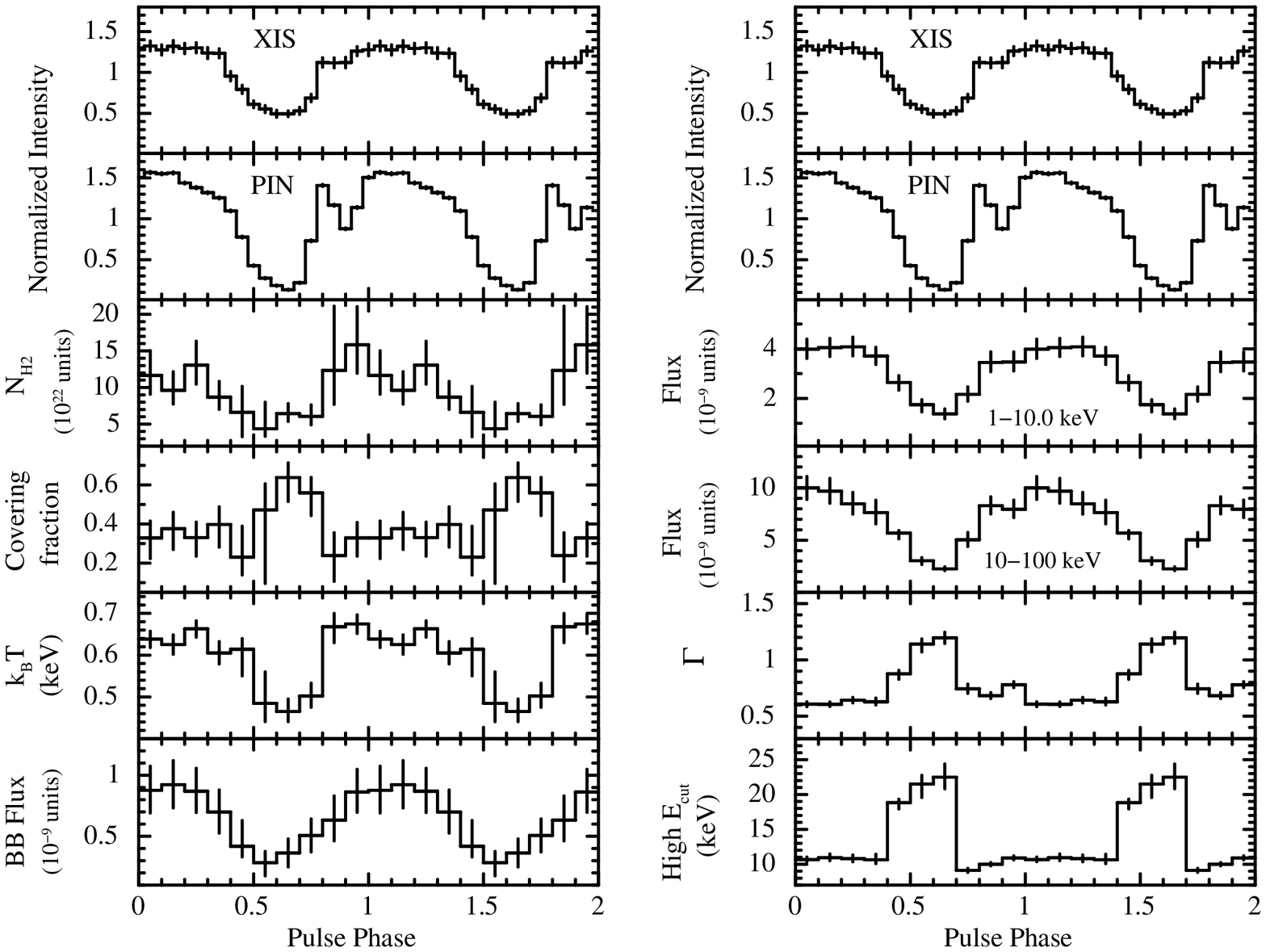}
\caption{Spectral parameters obtained from the phase-resolved spectroscopy 
of KS~1947+300 during second ${\it Suzaku}$ observation in 2013 November. 
The first and second panels in both sides show pulse profiles of the pulsar 
in 0.5-10~keV (XIS-0) and 10-70~keV (HXD/PIN) energy ranges. The values of 
$N_{H_2}$, covering fraction, blackbody temperature and blackbody flux
are shown in third, fourth, fifth and sixth panels in left side of the figure, 
respectively. The  soft X-ray flux in 1-10 keV range, 
hard X-ray flux in 10-100 keV range,  photon index and high energy cutoff 
are shown in third, fourth, fifth and sixth panels in right side 
of the figure, respectively. The blackbody flux and source 
fluxes in 1-10 and 10-100 keV are expressed in units of $10^{-9}$ ergs cm$^{-2}$ 
s$^{-1}$. The errors in the figure are estimated for 90\% confidence level.}
\label{pr2}
\end{figure*}

To investigate the nature of the absorption dip in hard X-ray pulse profiles, as well as the 
nature of the soft excess component and the evolution of other spectral parameters 
during both {\it Suzaku} observations, pulse-phase resolved spectroscopy was carried 
out by accumulating source spectra from XISs, PIN and GSO detectors in 9 and 10 phase 
bins for first and second observations, respectively. Background spectra, response 
matrices and effective area files used in phase-averaged spectroscopy were also used 
in the phase-resolved spectroscopy. Simultaneous spectral fitting was carried out for 
phase-resolved spectra obtained from both observations by using partial covering 
NPEX continuum model along with other components.  During fitting, the equivalent 
galactic hydrogen column density (N$_{H_1}$; expected to be constant along the source
direction), energy and width of iron emission lines, cyclotron line parameters 
(line energy, width and depth) and relative
instrument normalizations were fixed at corresponding phase-averaged values. Due to 
lack of sufficient photons, the iron emission lines were not resolved during the 
phase-resolved spectral fitting. It was found that the change in spectral parameters 
over pulse phase are similar for both the observations and are shown in Figures~\ref{pr1} 
\& ~\ref{pr2} for the first and second observation, respectively. Pulse profiles of the 
pulsar obtained from XIS-0 and PIN light curves of both observations are shown in top 
two panels of left and right panels of Figures~\ref{pr1} \& ~\ref{pr2}. Changes in 
the spectral parameters such as additional column density (N$_{H_2}$), covering 
fraction, blackbody temperature for soft X-ray excess and soft X-ray excess flux 
with pulse phases are shown in subsequent panels on the left sides of Figures~\ref{pr1} 
\& \ref{pr2}. The panels in right side of Figures~\ref{pr1} \& \ref{pr2} show the 
changes in soft (1-10 keV) and hard X-ray (10-100 keV) fluxes, power-law photon index 
and cutoff energy.  

\begin{figure}
\centering
\includegraphics[height=5.in, width=3.7in, angle=-90]{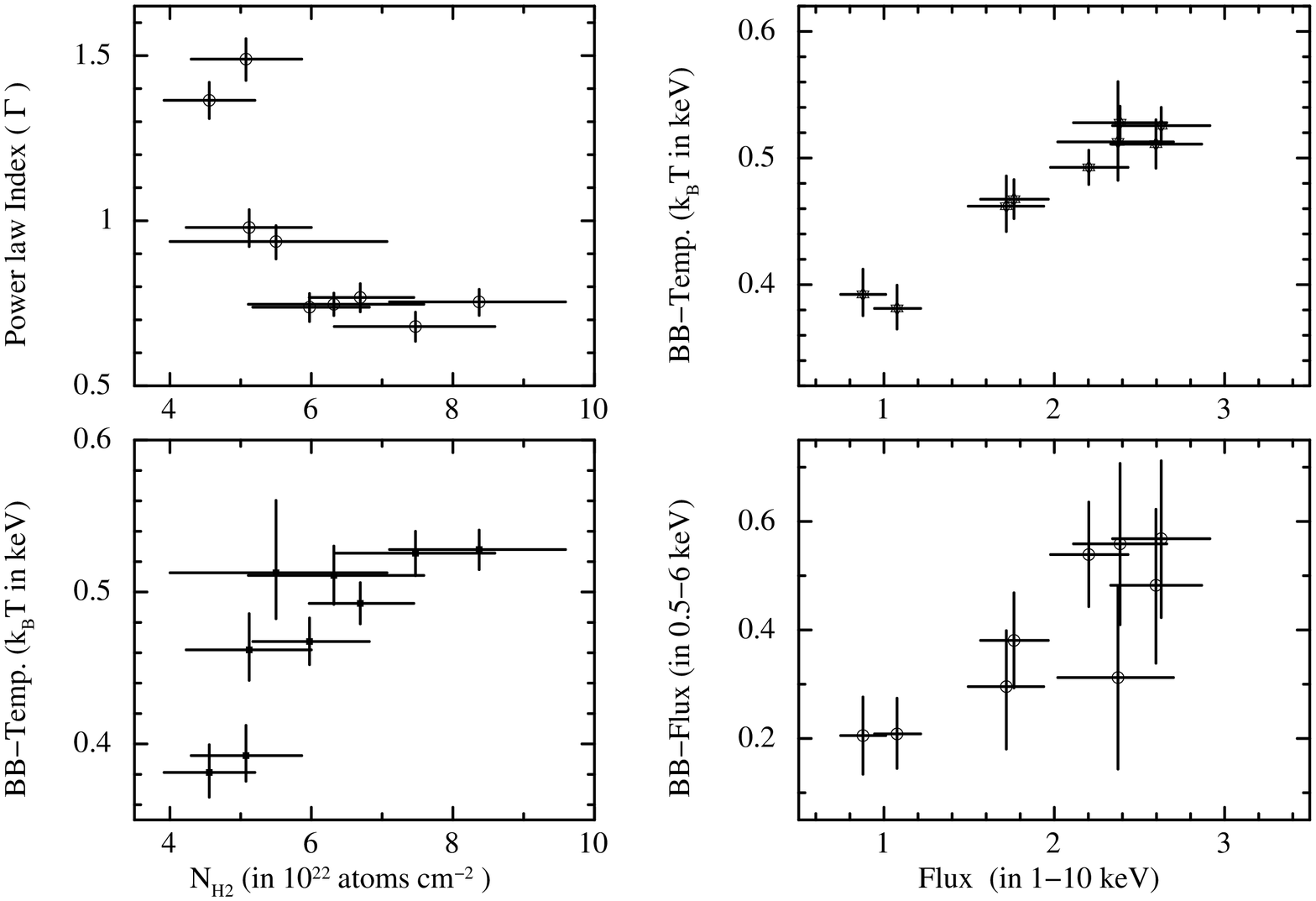}
\caption{Dependence of different spectral parameters obtained from the phase-resolved 
spectroscopy of KS~1947+300 during first {\it Suzaku} observation. The blackbody flux 
and 1-10 keV source flux are quoted in the units of $10^{-9}$ ergs cm$^{-2}$ s$^{-1}$.}
\label{cor1}
\end{figure}

\begin{figure}
\centering
\includegraphics[height=5.in, width=3.7in, angle=-90]{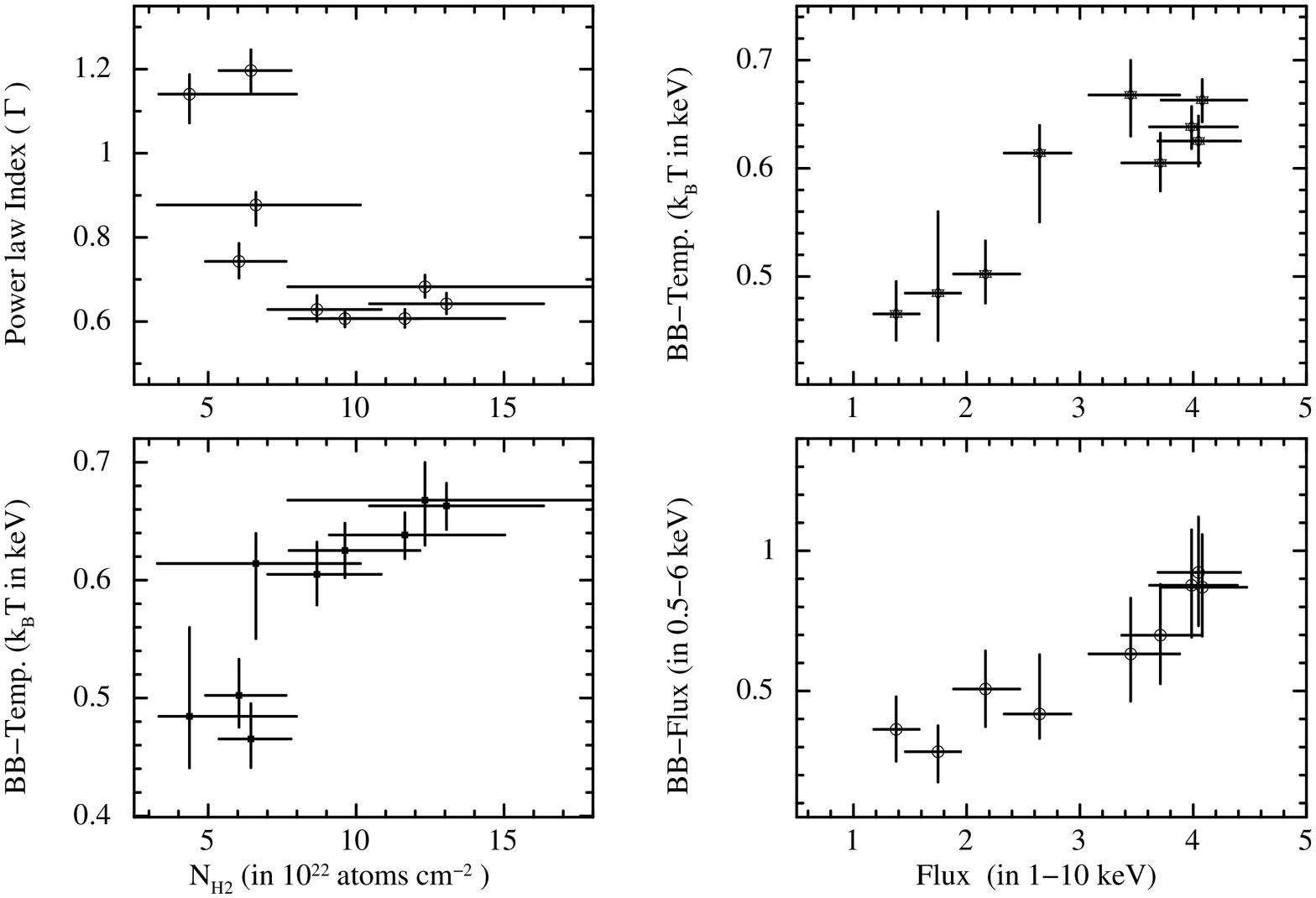}
\caption{Dependence of different spectral parameters obtained from the phase-resolved 
spectroscopy of KS~1947+300 during second {\it Suzaku} observation. The blackbody flux 
and 1-10 keV source flux are quoted in units of $10^{-9}$ ergs cm$^{-2}$ s$^{-1}$.}
\label{cor2}
\end{figure}

All the spectral parameters plotted in Figures~\ref{pr1} \& \ref{pr2} were found to
be variable with pulse phase of the pulsar. Additional column density was found to be
marginally higher at the phase of the absorption dip in hard X-ray pulse profile. 
Blackbody temperature, blackbody flux and source flux in the 1-10 keV range showed
similar variation pattern as the soft X-ray pulse profile over pulse phases. This 
confirmed the pulsating nature of the soft X-ray excess component in phase with the 
source flux. The values of power-law photon index and high energy cutoff were found 
higher during the main dip phases. 
Dependence of several spectral parameters on additional column density and source
flux in 1-10 keV range were investigated and are shown in Figures~\ref{cor1} \& 
\ref{cor2}. The left panels showed the dependence of power-law photon index and 
blackbody temperature on additional column density where as the right panels showed
the dependence of blackbody temperature and blackbody flux on source flux in 1-10 keV
range. It was found that the blackbody temperature and flux showed positive correlation 
with the soft X-ray flux in 1-10 kev range. The value of power-law photon index was found 
to be high at low value of additional column density (N$_{H_2}$) and decreased with the 
increase in N$_{H_2}$.

\begin{table*}
\centering

\caption{Best-fitting spectral parameters (with 90\% errors) obtained from two {\it Suzaku} 
observations of KS~1947+300 during 2013 October outburst. Model-1 : partial covering 
NPEX model with  Gaussian components and a cyclotron absorption line; 
Model-2 : partial covering high-energy cutoff model with  Gaussian components and 
a cyclotron absorption line; Model-3: partial covering cutoff power law model with
Gaussian components and a cyclotron absorption line. The cyclotron line parameters were fixed
at the values from F{\"u}rst et al. (2014).}
\begin{tabular}{lllllll}
\hline
\hline \\
Parameter                      &  \multicolumn{3}{c}{2013 October (Obs.1)}     	&   \multicolumn{3}{c}{2013 November (Obs.2)}   \\ 
\\
                                &Model-1             &Model-2            &Model-3           &Model-1		&Model-2		&Model-3        \\
\hline
\\
N$_{H1}$$^a$                    &0.50$\pm$0.02       &0.50$\pm$0.02      &0.52$\pm$0.02	    &0.48$\pm$0.02      &0.50$\pm$0.02		&0.53$\pm$0.02   \\
N$_{H2}$$^b$                    &7.6$\pm$1.0         &8.2$\pm$1.1        &7.6$\pm$1.1	    &11.3$\pm$2.5       &10.7$\pm$1.7		&12.1$\pm$4.8   \\
Covering fraction               &0.44$\pm$0.06       &0.45$\pm$0.06      &0.43$\pm$0.06	    &0.35$\pm$0.06	&0.35$\pm$0.06		&0.27$\pm$0.07   \\
Photon Index ($\Gamma$)         &0.67$\pm$0.03       &0.95$\pm$0.04      &0.92$\pm$0.05	    &0.62$\pm$0.04      &0.93$\pm$0.04		&0.93$\pm$0.03   \\
E$_{cut}$ (keV)	                &10.2$\pm$0.3        &5.4$\pm$0.4       &20.2$\pm$0.9      &10.6$\pm$0.4       &5.9$\pm$0.3		&21.6$\pm$0.7  \\
E$_{fold}$ (keV)		&--		     &21.0$\pm$0.7	 &--                &--			&21.6$\pm$0.7		&--		\\

Blackbody temp. kT (keV)	&0.54$\pm$0.02       &0.56$\pm$0.02	 &0.54$\pm$0.02	    &0.63$\pm$0.03	&0.65$\pm$0.02		&0.65$\pm$0.03    \\
Blackbody flux$^c$ 	        &0.88$\pm$0.13	     &0.96$\pm$0.13      &0.77$\pm$0.13     &1.34$\pm$0.19      &1.40$\pm$0.20          &1.02$\pm$0.19   \\

\\
{\it Emission lines } 
\\
Fe$~K\alpha$ line energy (keV)  &6.42$\pm$0.03       &6.42$\pm$0.03      &6.42$\pm$0.03     &6.45$\pm$0.02 	&6.45$\pm$0.02		&6.45$\pm$0.02   \\
Width of Fe line (keV)          &0.01$_{-0.01}^{+0.06}$	&0.01$_{-0.01}^{+0.06}$  &0.01$_{-0.01}^{+0.06}$      &0.01$_{-0.01}^{+0.04}$ &0.01$_{-0.01}^{+0.04}$  &0.01$_{-0.01}^{+0.04}$\\
Eq. width of Fe line (eV)       &18$\pm$2	     &19$\pm$2		 &19$\pm$2	    &20$\pm$3		&20$\pm$3		&22$\pm$2         \\
\\
 Line energy (keV)              &6.66$\pm$0.05       &6.66$\pm$0.05      &6.66$\pm$0.05     &6.71$\pm$0.04 	&6.71$\pm$0.04		&6.71$\pm$0.02   \\
 Line width (keV)               &0.01$_{-0.01}^{+0.07}$ &0.01$_{-0.01}^{+0.07}$  &0.01$_{-0.01}^{+0.07}$    &0.01$_{-0.01}^{+0.07}$ &0.01$_{-0.01}^{+0.07}$	&0.01$_{-0.01}^{+0.07}$	 \\
Equivalent width  (eV)          &17$\pm$2	     &18$\pm$2		 &18$\pm$2	    &11$\pm$3		&11$\pm$3		&13$\pm$3         \\
\\
Source flux  \\
Flux$^c$ (1-10 keV)   		&2.6$\pm$0.2       &2.7$\pm$0.2          &2.6$\pm$0.2        &4.3$\pm$0.4        &4.3$\pm$0.4            &4.1$\pm$0.4       \\
Flux$^c$ (10-70 keV)  		&5.4$\pm$0.5        &5.3$\pm$0.4         &5.3$\pm$0.3       &6.6$\pm$0.6        &6.5$\pm$0.5            &6.5$\pm$0.4       \\ 
Flux$^c$ (70-100 keV) 		&0.21$\pm$0.03      &0.27$\pm$0.02       &0.27$\pm$0.02     &0.38$\pm$0.03      &0.38$\pm$0.03          &0.38$\pm$0.02     \\
\\

Reduced $\chi^2$                 &1.18 (942)        &1.25 (970)          &1.25 (971)        &1.08 (970)         &1.10 (942)             &1.13 (971) \\
\\
\hline
\hline\\
\end{tabular}
\\
\flushleft
$^a$ : Equivalent hydrogen column density (in 10$^{22}$ atoms cm$^{-2}$ units).\\ 
$^b$ : Additional hydrogen column density (in 10$^{22}$ atoms cm$^{-2}$ units). \\
$^c$ : Absorption corrected  flux in units of 10$^{-9}$  ergs cm$^{-2}$ s$^{-1}$. \\
\label{spec_par}
\end{table*}


\section{Discussion and Conclusions}

\subsection{Spin-period and magnetic field of the pulsar}

KS~1947+300 was observed with {\it Suzaku} at two epochs during its 2013 X-ray 
outburst. Though the observations were only a month apart, estimated spin periods 
of the pulsar during both the observations were found to be different. The spin
period during the second observation was smaller than the first observation, 
indicating the pulsar was spinning up during the X-ray outburst. While comparing 
the recent measurements of spin period of the pulsar with several observatories 
(as quoted in Table~1), it was found that the pulsar was continuously spinning 
up during whole of the 2013 X-ray outburst. During X-ray outbursts, spinning-up 
of the neutron star is expected due to transfer of angular momentum from 
accreting matter at the magnetic poles. Ghosh \& Lamb (1979) formulated the
dependence of spin-up rate of a pulsar on its luminosity as $\dot{P}\propto{L^{6/7}}$. 
In the present work, the pulsar spin period was found to 18.7878~s ({\it Suzaku}) 
at the peak of the outburst which decreased to 18.77088~s ({\it NuSTAR} observation) 
during the decay of the outburst (see Table~\ref{pulse}). The observed spin-up of 
KS~1947+300 can be attributed to the change in angular momentum due to the torque 
exerted by accreting matter on the neutron star. A similar type of rapid 
spin-up was also observed during the declining phase of the 2001 outburst of the pulsar 
(Naik et al. 2006).

 The pulsar was showing a spin-up trend during {\it Suzaku} and {\it NuSTAR} 
observations. The observed spin-up (angular frequency) rate ($\dot\omega_{su}$) 
can be used to estimate the magnetic field of the pulsar by considering the 
quasi-spherical settling accretion theory (Shakura et al. 2012; Postnov et al. 
2015). According to this theory, 

\begin{equation}\label{1}
 \dot\omega_{su}\simeq 10^{-9}[Hz~d^{-1}]~\Pi_{su}~\mu_{30}^{1/11}
v_{8}^{-4}\left(\frac{P_{b}}{10~d}\right)^{-1}\dot M_{16}^{7/11} 
\end{equation}   

where $\dot\omega_{su}$ is spin-up rate which is estimated to be 
1.27$\times$10$^{-5}$[Hz~d$^{-1}$] (present case) and $\Pi_{su}$ is the 
dimensionless parameter of the theory of setting accretion. The value of 
$\Pi_{su}$ is independent of the system and is in the range of $\sim$4.6 to 10 
(Shakura et al. 2012; Postnov et al. 2015). In the present case, $\Pi_{su}$ was 
assumed to be 4.6. The dipole magnetic moment of the neutron star
$\mu_{30}$=$\mu$/10$^{30}$[G~cm$^3$] and related to the magnetic field (B) 
through the relation $\mu$=BR$^3$/2 (R is the neutron star radius, assumed 
to be 10~km). Stellar wind velocity v$_8$=v/10$^8$[cm~s$^{-1}$] is considered 
to be 200 km~s$^{-1}$ for typical Be/X-ray binaries (Waters et al. 1988). 
The mass accretion rate $\dot M_{16}$=$\dot M$/10$^{16}$[g~s$^{-1}$]  
for the luminosity of 10$^{38}$~erg~s$^{-1}$ was estimated to be 
$\dot M_{16}$=100 during {\it Suzaku} observations. The orbital period 
P$_b$ of KS~1947+300 is 40.42~d (Galloway et al. 2004). Using above parameters,
the magnetic field of the pulsar was estimated to be $\sim$1.2$\times$10$^{12}$~G.
The estimated value of magnetic field by using observed spin-up rate in 
KS~1947+300 was found to agree with that obtained from the detection of cyclotron 
absorption line at 12.2~keV.

\subsection{Pulse profiles}

In the present work, the pulse profile of KS~1947+300 was found to be simple at 
lower energies. As the energy increases, a dip like structure appears  in 
the pulse profile and is detected up to 70~keV. The depth of the dip is found to increase with 
the energy and found maximum in 30-40~keV range. Such type of behavior is not 
generally seen in pulse profiles of other Be/X-ray binary pulsars. We tried to
investigate the cause of the absorption dip in pulse profiles at
hard X-rays ($>$10 keV) through phase-resolved spectroscopy. A marginal 
enhancement in the additional column density at dip phase was detected. 
 Such low value ($\le$20$\times$10$^{22}$cm$^{-2}$) of additional 
column density, however, can not absorb the hard X-ray photons up to $\sim$70~keV. 
KS~1947+300 was also observed at different luminosity levels with several observatories
such as {\it BeppoSAX}, {\it RXTE}, {\it INTEGRAL} and {\it NuSTAR}. The pulse profiles 
obtained from these observations were found to be similar to that obtained from 
{\it Suzaku} observations. The dip was only seen in hard X-rays (Galloway et al. 2004; 
Tsygankov et al. 2005; Naik et al. 2006; F{\"u}rst et al. 2014). Therefore, the presence 
of the dip in hard X-ray pulse profiles of KS~1947+300 is possibly intrinsic to the 
pulsar. 

In general, the pulse profiles of the Be/X-ray binary pulsars are seen to be 
strongly energy and luminosity dependent. Single or multiple absorption dips, 
prominent at soft X-ray, are seen in the pulse profiles of these pulsars (Paul 
\& Naik 2011 and references therein). With increase in energy, the depth of the 
dip decreases and becomes invisible at higher energies. It is widely believed 
that these dips in the pulse profile are due to obscuration/absorption of soft 
X-ray photons by matter present close to the neutron star. In some cases, 
single or multiple dips were observed at high energies e.g. up to 70~keV in 
pulse profiles of EXO~2030+275 (Naik et al. 2013; Naik et al. 2015). 
The presence of additional dense matter at dip phases was detected from 
phase-resolved spectroscopy and was interpreted as the cause of 
absorption dips in the pulse profiles of EXO~2030+375. In KS~1947+300 
(present work), however, the origin of the dip in hard X-ray pulse profiles 
(up to $\sim$70~keV) is not due to the presence of additional matter at certain
phase of the pulsar. 

The pulse profile of the pulsars can be affected by the cyclotron resonance 
scattering and geometrical effects. These effects can play a vital role and 
shape anomaly or dip in the pulse profiles. In KS~1947+300, cyclotron absorption 
line was detected at $\sim$12.2~keV (F{\"u}rst et al. 2014). 
The beam function of an accreting pulsar can be affected by the presence of strong 
cyclotron resonance scattering which can produce a significant change in the pulse 
profile e.g. phase-shift (lag) (Sch{\"o}nherr et al. 2014). Similar effects were detected
in Be/X-ray binary pulsars such as V~0332+53 (Tsygankov et al. 2006), 
4U~0115+63 (Ferrigno et al. 2011) and GX~304-1 (Jaisawal et al. 2016).
However, this is not the case in KS~1947+300 as the strength of the observed 
dip increased with energy and became prominent in the $\sim$30-40~keV energy range.
Around this energy, however, the effects of cyclotron resonance scattering is
not be as effective as compared to energies closer to $\sim$12~keV. In addition, 
any significant change in the pulse profiles (beam pattern) or phase-lags was 
not observed in the energy resolved pulse profiles (Fig.~\ref{er1} \& \ref{er2}). 
Therefore, we expect that the cyclotron scattering is not causing the hard 
X-ray dip in the pulse profiles of KS~1947+300. Alternatively, the presence of 
single dip in hard X-ray profiles can suggest a way of direct viewing 
the accretion column along the magnetic axis of the pulsar. At such high 
luminosity ($\sim$10$^{38}$erg~s$^{-1}$) like the one of the {\it Suzaku} 
observations of KS~1947+300, a radiation pressure dominated shock is expected 
to form above the neutron surface which can absorb the photons up to higher 
energies. In this case, the position of the absorption dip should be at the 
peak of the pulse profiles. However, the asymmetric phase position of the dip 
with respect to main dip in pulse profiles (Fig.~\ref{er1} \& \ref{er2}) discards 
the hypothesis of direct viewing of the pulsar along the magnetic axis. 
It is accepted that the pulse profile of X-ray pulsars depends on the 
geometry and viewing angle of the emission region or accretion column 
(Kraus et al. 1995; Caballero et al. 2011; Sasaki et al. 2012). 
We suggest that the dip in the hard X-ray pulse profiles of KS~1947+300 
is due to these geometrical effects. The dip was absent in the soft X-ray 
pulse profiles. The presence of strong soft X-ray excess (which was found 
pulsating in phase with the neutron star) may cancel the effect of the 
absorption dip, producing single pulse profiles in soft X-ray bands.

\subsection{Spectroscopy}

In this paper, broad-band phase-averaged and phase-resolved spectra of KS~1947+300 
are presented by using two {\it Suzaku} observations of the 2013 giant outburst. 
During both observations, the values of estimated galactic column density were 
comparable. However, the values of the additional column density were found to be 
significantly higher than the galactic value. The higher values of additional 
absorption column density indicates the presence of additional matter near the 
neutron star during the X-ray outburst. During both the observations, a soft 
X-ray excess was clearly detected and its temperature was found to be high at 
the peak of the outburst (second observation). Assuming the blackbody 
emitting region to be spherically symmetric, the radius of soft X-ray excess 
emitting region in KS~1947+300 is estimated to be  $\sim$29--31~km. It implies 
that the soft X-ray excess emitting region is close to the neutron star surface. 
The pulsating nature of the soft X-ray excess in KS~1947+300 agrees with the above 
argument. Therefore, the accretion column and/or accretion streams are the most 
probable origin site of soft X-ray excess emission in KS~1947+300 (Naik \& Paul 
2002; Naik \& Paul 2004; Hickox et al. 2004). 

Apart from the detection of soft X-ray excess and presence of additional 
matter around the pulsar, change in power-law photon index and high energy 
cutoff with pulse phase was seen in KS~1947+300. Similar type of variation 
was also seen in other Be/X-ray binary pulsar such as EXO~2030+275 (Naik et al. 
2013; Naik \& Jaisawal 2015). Narrow iron K$_\alpha$ and He-like iron 
emission lines at $\sim$6.4 and 6.7~keV were detected during 
both the observations. During {\it Beppo}SAX observations of KS~1947+300 
in 2001 outburst, an emission line at 6.7~keV was detected whereas iron 
K$_\alpha$ line was absent in the pulsar spectra (Naik et al. 2006). The 
6.7~keV line was identified as the emission feature from helium like iron 
atoms. A cyclotron absorption feature at $\sim$12.2~keV was detected in 
KS~1947+300 from {\it NuSTAR} observations during 2013 outburst (F{\"u}rst 
et al. 2014). Detection of cyclotron line is an unique tool to direct 
estimate the magnetic field of the pulsar  by using 12-B-12 rule or 
{\it E$_{cyc}$=11.6B$_{12}\times(1+z){^{-1}}$}. Using the detected cyclotron
absorption line at 12.2 keV, the strength of surface magnetic was estimated to be
$\sim$1.1$\times10^{12}$(1+z)~G (F{\"u}rst et al. 2014). Though the pulsar
was observed with {\it NuSTAR} at three epochs, the cyclotron line was detected 
only during the second observation and there was no signature of the presence of
its harmonics in the pulsar spectrum. We also did not find harmonics of the 12.2 keV
cyclotron line in pulsar spectra obtained from {\it Suzaku} observations. There 
are several pulsars where fundamental cyclotron line is seen in the broad-band 
spectra without the detection of its harmonics e.g. Cen~X-3 (Suchy et al. 2008; 
Naik et al. 2011), Swift~J1626.6-5156 (DeCesar et al. 2013), IGR~J17544-2619 
(Bhalerao et al. 2015). The 1-100~keV luminosity of the pulsar was estimated 
to be $\sim$9.8$\times$10$^{37}$ and 1.3$\times$10$^{38}$~ergs s$^{-1}$ 
during first and second {\it Suzaku} observations, respectively. Critical 
luminosity was calculated to investigate the luminosity regime of the pulsar 
by assuming parameters of a canonical neutron star with cyclotron line energy 
at 12.2~keV in the relation of Becker et. al. (2012). This was estimated to be
$\sim$1.6$\times$10$^{37}$~ergs s$^{-1}$. It is clear that the pulsar was 
accreting in the super-Eddington regime (above the critical luminosity) during 
2013 October (present work) and 2000 November (Naik et al. 2006) outbursts.  

In summary, we reported on the timing and broad-band spectral properties of 
the pulsar KS~1947+300 by using {\it Suzaku} observations taken during the 2013 outburst. 
Soft X-ray pulse profiles were found to be smooth and single peaked. However, 
hard X-ray pulse profiles showed the presence of an absorption dip like 
feature. The 1-110~keV broad-band spectrum of the pulsar was described with a 
partially absorbed NPEX continuum model along with a blackbody component. 
Phase-resolved spectroscopy revealed marginal enhancement in the additional 
column density at dip phase which suggests that the dip is not because 
of the absorption of hard X-ray photons. Other mechanism such as geometrical 
effect can be the probable cause for the presence of dip in the hard X-ray 
pulse profiles of KS~1947+300. Detection of pulsation in the soft X-ray 
excess flux confirmed that the emitting region is close to the neutron
star e.g. near the accretion column. The presence 
of soft X-ray excess may be the cause of the absence of the dip in soft X-ray 
profiles. We estimated the magnetic field of the pulsar by using the observed
spin-up rate during {\it Suzaku} and {\it NuSTAR} observations. The value was found 
to be 1.2$\times$10$^{12}$~G and comparable to that obtained from the cyclotron 
line energy.

\section*{Acknowledgments}
We sincerely thank the referee for his valuable comments and suggestions which 
improved the paper significantly. The research work at Physical Research 
Laboratory is funded by the Department of Space, Government of India. 
The authors would like to thank all the members of the {\it Suzaku} for their 
contributions in the instrument preparation, spacecraft operation, software 
development, and in-orbit instrumental calibration. This 
research has made use of data obtained through HEASARC Online Service, provided 
by the NASA/GSFC, in support of NASA High Energy Astrophysics Programs.

\label{lastpage}

\begin{thebibliography}{}
\bibitem[]{}Becker, P. A., Klochkov, D., Sch{\"o}nherr, G., Nishimura, O., Ferrigno, C. et al. 2012, A\&A, 544, 123
\bibitem[]{}Bhalerao, V., et al. 2015, MNRAS, 447, 2274
\bibitem[]{}Borozdin, K., Gilfanov, M., Sunyaev, R., et al. 1990, SvAL, 16, 345 
\bibitem[]{}Caballero, I., Kraus, U., Santangelo, A., Sasaki, M., Kretschmar, P., 2011, A\&A, 526, 131C
\bibitem[]{}Caballero, I. \& Wilms, J., 2012, MmSAI, 83, 230
\bibitem[]{}Chakrabarty, D., Koh, T., Bildsten, L., et al. 1995, ApJ, 446, 826 
\bibitem[]{}DeCesar, M. E., Boyd, P. T., Pottschmidt, K., Wilms, J., Suchy, S., Miller, M. C., 2013, ApJ, 762, 61
\bibitem[]{}Ferrigno, C., Falanga, M., Bozzo, E., Becker, P. A., Klochkov, D., Santangelo, A. 2011, A\&A, 532, A76
\bibitem[]{}Finger, M. H., Jenke, P. A. and Wilson-Hodge, C. A. 2015, Astron. Telegram, 7017, 1
\bibitem[]{}F{\"u}rst, F., et al. 2014, ApJ, 784, L40 
\bibitem[]{}Galloway, D.~K., Morgan, E.~H., Levine A.~M. 2004, ApJ, 613, 1164 
\bibitem[]{}Ghosh, P., Lamb, F.~K. 1979, ApJ, 234, 296 
\bibitem[]{}Hickox, R. C., Narayan, R. \& Kallman, T. R. 2004, ApJ, 614, 881
\bibitem[]{}James, M., Paul, B., Devasia, J., Indulekha, K. 2010, MNRAS, 407, 285
\bibitem[]{}Jaisawal, G. K., Naik, S. \& Epili, P. 2016, MNRAS, (in press), arXiv:1601.02348
\bibitem[]{}Koyama, K. et al. 2007, PASJ, 59, 23
\bibitem[]{}Kraus, U., Nollert, H.-P., Ruder, H., Riffert, H. 1995, ApJ, 450, 763 
\bibitem[]{}Levine, A., Corbet, R. 2000, IAU Circ. 7523
\bibitem[]{}Mitsuda, K. et al., 2007, PASJ, 59, 1
\bibitem[]{}Naik, S. \& Paul, B. 2002, JApA, 23, 27
\bibitem[]{}Naik, S. \& Paul, B. 2004, ApJ, 600, 351
\bibitem[]{}Naik, S., Callanan, P.~J., Paul, B., Dotani, T. 2006, ApJ, 647, 1293 
\bibitem[]{}Naik, S., Paul, B., Ali, Z. 2011, ApJ, 737, 79 
\bibitem[]{}Naik, S., Maitra, C., Jaisawal, G. K., Paul, B. 2013, ApJ, 764, 158
\bibitem[]{}Naik, S. \& Jaisawal, G. K. 2015, RAA, 15, 537
\bibitem[]{}Negueruela, I., Israel, G.~L., Marco, A., Norton, A.~J., \& Speziali, R.  2003, A\&A, 397, 739
\bibitem[]{}Nowak, M. A., Hanke, M., Trowbridge, S. N., et al. 2011, ApJ, 728, 13
\bibitem[]{}Okazaki, A. T. \& Negueruela, I. 2001, A\&A, 377, 161
\bibitem[]{}Paul, B. \& Naik, S. 2011, Bull. Astron. Soc. India, 39, 429
\bibitem[]{}Postnov, K. A., Mironov, A. I., Lutovinov, A. A., Shakura, N. I., Kochetkova, A. Yu., Tsygankov, S. S. et al. 2015, MNRAS, 446, 1013
\bibitem[]{}Reig, P. 2011, Ap\&SS, 332, 1
\bibitem[]{}Sasaki, M., M{\"u}ller, D., Kraus, U., Ferrigno, C., Santangelo, A., 2012, A\&A, 540A, 35S
\bibitem[]{}Sch{\"o}nherr, G., et al. 2014, A\&A, 564, 8
\bibitem[]{}Serlemitsos, P. J., et al.  2007, PASJ, 59, S9
\bibitem[]{}Shakura, N., Postnov, K., Kochetkova, A., Hjalmarsdotter, L., MNRAS, 2012, 420, 216
\bibitem[]{}Skinner, G. K. 1989, IAU Circ. 4850
\bibitem[]{}Suchy, S., Pottschmidt, K., Wilms, J., Kreykenbohm, I., Sch{\"o}herr, G. et al. 2008, ApJ, 675, 1487
\bibitem[]{}Swank, J. \& Morgan, E. 2000, IAUC, 7531, 4 
\bibitem[]{}Takahashi, T., et al., 2007, PASJ, 59, 35
\bibitem[]{}Tsygankov, S.~S., Lutovinov, A.~A. 2005, AstL, 31, 88
\bibitem[]{}Tsygankov, S. S., Lutovinov, A. A., Churazov, E. M. \& Sunyaev, R. A. 2006, MNRAS, 371, 19
\bibitem[]{}Waters, L. B. F. M., van den Heuvel, E. P. J., Taylor, A. R., Habets, G. M. H. J., Persi, P. 1988, A\&, 198, 200






\end{thebibliography}
\end{document}